\documentclass[12pt,a4paper]{article}
\usepackage{amssymb}
\usepackage{amsmath}

\newcounter{mnotecount}[section]

\newcommand{\mnote}[1]
{\protect{\stepcounter{mnotecount}}$^{\mbox{\footnotesize
$
\bullet$\themnotecount}}$ \marginpar{
\raggedright\tiny\em
$\!\!\!\!\!\!\,\bullet$\themnotecount: #1} }

\newcommand{\koniec}{\begin{flushright}$\Box$\end{flushright}}
\newcommand\qed{\phantom{\underline{y}}\hfill\hfill$\square$}
\newtheorem{theo}{Theorem}[section]
\newtheorem{prop}[theo]{Proposition}

\newcommand{\Dd}{{\mathcal D}}
\newcommand{\R}{\mathbb{R}}

\newcommand\p{\partial}
\newcommand\be{\begin{equation}}
\newcommand\ee{\end{equation}}

\newcommand{\RP}{\mathbb RP}

\newtheorem{theorem}{Theorem}

\begin{document}

\title{On the Einstein-Weyl and  conformal self-duality equations}

\author{M. Dunajski, E.V. Ferapontov and B. Kruglikov}
     \date{}
     \maketitle
     \vspace{-5mm}
\begin{center}
DAMTP,  Centre for Mathematical Sciences,\\
University of Cambridge,\\
Wilberforce Road, Cambridge CB3 0WA, United Kingdom;\\
and\\
Department of Computer Science,\\
Faculty of Physics and Applied Informatics,\\
University of Lodz,\\
Pomorska 149/153, 90-236 Lodz, Poland;\\
\ \\
Department of Mathematical Sciences \\ Loughborough University, Loughborough, \\
Leicestershire LE11 3TU, United Kingdom; \\
  \ \\
Institute of Mathematics and Statistics, NT-Faculty, \\
University of Troms\o, Troms\o\ 90-37, Norway\\
[2ex]
e-mails: \\[1ex] \texttt{M.Dunajski@damtp.cam.ac.uk}\\
\texttt{E.V.Ferapontov@lboro.ac.uk}\\
\texttt{boris.kruglikov@uit.no} \\

\end{center}

\vspace{1cm}
\begin{abstract}

The equations governing anti-self-dual and Einstein-Weyl conformal geometries can be regarded as
`master dispersionless systems' in four and three dimensions respectively. Their integrability by twistor methods
has been established by Penrose and Hitchin. In this  note we present, in  specially adapted coordinate systems, explicit forms of the corresponding equations
and their Lax pairs. In particular, we demonstrate that any Lorentzian Einstein-Weyl structure is locally given by a 
solution to the Manakov-Santini system, and we find a system of two coupled third-order scalar PDEs for a general anti-self-dual conformal structure
in neutral signature. 


\end{abstract}

\newpage

\section{Introduction}

There exist two key `integrable' conformal geometries, namely  Einstein-Weyl geometry in three dimensions, and  anti-self-dual (ASD) geometry in four dimensions (see \cite{Calderbank} for a comprehensive
overview). 
In spite of their fundamental role  in twistor theory, general relativity and the theory of dispersionless integrable systems, these geometries remain largely unknown to the integrable systems community due to a lack of explicit coordinate formulae for the underlying PDEs and Lax pairs. The aim of this paper is to present them in their simplest possible  forms, in specially adapted coordinates.

In Section 2 we discuss Einstein-Weyl (EW) structures  in three dimensions. Recall that an EW geometry on a three-dimensional manifold $M^3$ consists of a conformal structure $[g]$
and a symmetric connection
${\mathbb{D}}$ compatible with $[g]$ in the sense that, 
for any $g\in [g]$, \ 
\[
{\mathbb {D}}g=\omega\otimes g
\]
for some covector $\omega$, and such that
the trace-free part of the symmetrized Ricci tensor of ${\mathbb{D}}$
vanishes. Using Cartan's approach relating EW structures to a special class of third-order ODEs  we shall demonstrate the following
\begin{theorem}
\label{theo_ms}
There exists a local coordinate system $(x, y, t)$ on $M^3$ such that
any Lorentzian Einstein--Weyl structure is locally of the form
\begin{equation}
g = -(dy - v_x dt )^2 +4(dx - (u - v_y ) dt ) dt, ~~~
\omega= -v_{xx} dy+(4u_x -2v_{xy} +v_xv_{xx})dt,
\label{metricMS}
\end{equation}
where the functions  $u$ and $v$ on $M^3$ satisfy a coupled system of second-order PDEs
 \begin{equation}
P(u)+u_x^2=0, ~~~ P(v)=0,
\label{MS}
\end{equation}
where 
$$
P=\p_x\p_t-\p_y^2+(u-v_y){\p_x}^2+v_x{\p_x\p_y}. 
$$
\end{theorem}
System (\ref{MS}) is known as the Manakov-Santini system, and was originally
derived in \cite{Man-San}
as a two-component generalization of the dispersionless KP equation.
It was shown in \cite{Dun3} that any solution to 
(\ref{MS}) gives rise to an EW structure of the form (\ref{metricMS}),
but the question whether all EW structures arise in that way has remained 
open. System (\ref{MS}) possesses the Lax representation  $[X_1, X_2]=0$ where
$$
X_1=\p_y-(\lambda+v_x)\p_x-u_x\p_\lambda, \quad X_2=\p_t-(\lambda^2+v_x\lambda-u+v_y)\p_x-(u_x\lambda+u_y)\p_\lambda
$$
are vector fields on the correspondence space $M^3\times\RP^1$,
where $\lambda\in\RP^1$.
Projecting  integral surfaces of the distribution spanned by $X_1, X_2$ from $M\times\RP^1$ to $M^3$ yields a two-parameter family of surfaces in $M^3$ 
which are null with respect to the conformal structure $[g]$, and totally geodesic in the  Weyl connection 
$\mathbb{D}$ (the existence of such surfaces 
is equivalent to the EW property \cite{Cartan}).
System (\ref{MS})  consists of 2 second-order PDEs for 2 functions of 3 independent variables, and
its general solution in real-analytic category depends on 4 arbitrary functions of 2 variables: this confirms  Cartan's count  \cite{Cartan}.

The relation between  EW geometry and dispersionless integrable systems in three dimensions has been known since \cite{Ward, Dun4, Dun3}. It was observed recently in \cite{FerKrug} that the dispersionless integrability of various classes of second-order PDEs is equivalent to the EW property of conformal structures defined by their principal  symbols. Moreover, in many examples the covector $\omega$ is expressed in terms of  $g\in [g]$ by the  universal explicit formula
\begin{equation}
\omega_k=2g_{kj}\mathcal{D}_{x^s}(g^{js})+\mathcal{D}_{x^k}(\ln\det g_{ij}).
\label{omega}
\end{equation}
Here $\mathcal{D}_{x^s}$ denotes total derivative with respect to  $x^s$, and $(x^1, x^2, x^3)=(x, y, t)$. In three dimensions this formula  is  invariant
under the transformation 
\be
\label{gauge_ew}
g\to \varphi^2 g, \quad \omega \to \omega + 2d\ln\varphi, \quad
\mbox{where}\quad \varphi:M^3\rightarrow \R^+,
\ee
that keeps the Einstein-Weyl equations invariant. The Manakov-Santini system  fits into this framework: the covector in (\ref{metricMS}) is  given by formula (\ref{omega}), and the principal symbol of system (\ref{MS}) equals $P^2$ (it is doubly degenerate) where $P$, viewed as a symmetric bivector, gives rise to the EW metric $g$ given by (\ref{metricMS}).

\medskip

In Section 3 we study ASD conformal structures in four dimensions. Recall that a conformal structure $[g]$  is called anti-self-dual
if the self-dual  part of the Weyl tensor of any $g\in[g]$ vanishes: $W_+=\frac12(W+*W)=0$. We shall establish the following
\begin{theorem}
\label{prop_ASD}
There exist local coordinates $(w, z, x, y)$ such that any ASD conformal structure
in signature $(2, 2)$  is locally represented by a metric 
\be
\label{ASDmetric1}
g=dwdx+dzdy+F_ydw^2-(F_x+G_y)dwdz+G_xdz^2,
\ee
where the functions
$F, G:M^4\rightarrow \R$ satisfy a coupled system of third-order PDEs,
\be
\label{sd_3rd}
\p_x(Q(F))-\p_y(Q(G))=0, \quad
(\p_w-F_y\p_x+G_y\p_y)Q(G)+(\p_z+F_x\p_x-G_x\p_y)Q(F)=0,
\ee
where
\[
Q=\p_w\p_x+\p_z\p_y-F_y{\p_x}^2-G_x{\p_y}^2+(F_x+G_y)\p_x\p_y.
\]
\end{theorem}
System (\ref{sd_3rd}) arises as $[X_1, X_2]=0$ from the dispersionless
Lax pair
\be
\label{new_lax}
X_1=\p_w-F_y\p_x+G_y\p_y+\lambda\p_y+Q(F)\p_\lambda, \quad X_2=\p_z+F_x\p_x-G_x\p_y+\lambda\p_x-Q(G)\p_\lambda.
\ee
Projecting  integral surfaces of the distribution spanned by $X_1, X_2$ in the correspondence space $M^4\times\RP^1$ to $M^4$ gives a 
3-parameter family of totally null surfaces with self-dual tangent bi-vector. These are the $\alpha$-surfaces of the corresponding conformal structure $[g]$. The existence of such surfaces is equivalent to the ASD property \cite{Penrose}. 
System (\ref{sd_3rd}) consists of 2 third-order PDEs for 2 functions of 4 independent variables, and its general solution depends on 6 arbitrary functions of 3 variables: this agrees with 
the count of  \cite{Grossman} based on the  Cartan-K\"ahler theory.
The Lax pair  (\ref{new_lax}) has appeared in \cite{Bog_0}, where the Riemann--Hilbert
problem associated to (\ref{new_lax}) has been formulated. The fact that the system
of 3 second order PDEs derived in \cite{Bog_0, Bog_1} leads to system (\ref{sd_3rd}) has been recently  pointed out in \cite{yi}.

ASD equations and their reductions provide a number of key examples of dispersionless integrable systems in four dimensions \cite{Plebanski, Dun5, DP}. It was conjectured in  \cite{FerKrug} that the dispersionless integrability of some four-dimensional PDEs  is equivalent to the requirement that the principal symbol of the equation defines a conformal structure that must be ASD on every solution. In \cite{FerKrug} this was demonstrated to be the case for integrable symplectic Monge-Amp\`ere equations \cite{DF}. The example of ASD equations fits into this scheme:  the principal symbol of system (\ref{sd_3rd}) equals $Q^3$ (it is triply degenerate), and the symmetric bivector $Q$ gives rise to the ASD metric $g$ given by (\ref{ASDmetric1}).

\noindent

\smallskip

{\bf Reality conditions.} If all coordinates and functions in 
Theorems \ref{theo_ms} and \ref{prop_ASD} are assumed to be real, then the corresponding 
conformal structures in three and four dimensions
have Lorentizan or $(2, 2)$ (also called neutral or Kleinian) signatures,
respectively. Alternatively, one can assume real analyticity and work in the complexified settings where all structures are assumed to be holomorphic. We shall make this additional assumption
whenever we rely on the Cauchy--Kowalevskaya theorem to 
assert that a general solution depends on $m$ functions
of $n$ variables.

\subsubsection*{Acknowledgements}
We are indebted to Leonid Bogdanov, Paolo Santini, and an anonymous referee for pointing out that the Lax formulation (\ref{new_lax}) 
first appeared in \cite{Bog_0}.

\section{Einstein-Weyl geometry}


The twistor integrability of  EW equations was established  in \cite{Hitchin}.
It was demonstrated in \cite{Dun4}  that   EW equations  possess a Lax pair given by two vector fields that form an integrable 
distribution, and may contain derivatives with respect to the spectral parameter.
Integral manifolds of this distribution provide  the  2-parameter family of null totally geodesic surfaces, and, as
shown by Cartan \cite{Cartan}, the EW property is equivalent to the existence of such family.
However, the explicit coordinate form of the Lax pair has not been exhibited in the general case. Below we list  various forms of  EW equations, as well as their Lax pairs, in specially adapted coordinates.


\subsection{Einstein-Weyl equations in  Cartan's approach}
Our proof of Theorem \ref{theo_ms} builds on Cartan's approach to Einstein-Weyl geometry
via special third-order ODEs \cite{Cartan1}. We shall briefly review it following the paper 
of Tod \cite{Tod}.
Consider an equivalence class of third-order ODEs
\be
\label{3rd_ode}
Y'''=F(X,Y,Y',Y''),
\ee
modulo point transformations 
$(X, Y)\rightarrow (\bar{X}(X, Y), \bar{Y}(X, Y))$. Here $'=d/dX$.
Let the general solution of (\ref{3rd_ode}) be of the form
\be
\label{EWsurface}
Y=Z(X, x^j),
\ee
where $x^j$ are coordinates on  the 
three--dimensional solution space $M^3$.
The necessary and sufficient conditions for the solution space
$M^3$ to carry an Einstein--Weyl structure such that the 2-parameter
family of surfaces in $M^3$ corresponding to fixing $(X, Y)$ in 
(\ref{EWsurface}) is null  and totally geodesic, are given by
the vanishing of the Wunshmann and Cartan invariants $W$ and $C$.
These invariants are given by
 $$
W=\tfrac16\Dd^2F_Q-\tfrac13F_Q\Dd F_Q-\tfrac12\Dd F_P+\tfrac2{27}F_Q^3+\tfrac13F_QF_P+F_Y,
 $$
 $$
C=(\tfrac13\Dd F_Q-\tfrac19F_Q^2-F_P)F_{QQ}+\tfrac23F_QF_{QP}-2F_{QY}+F_{PP}+2W_Q,
 $$
where  $\Dd=\partial_X+P\partial_Y+Q\partial_P+F\partial_Q$ is the total derivative.

The above $W$ is actually a relative contact invariant, while $C$ is a relative point invariant
(so their vanishing is an invariant condition for respective pseudogroups).


Following the approach of Tod \cite{Tod} (see 
\cite{Nurowski, Krynski} for other approaches),
the conformal structure and the covector are given by
\begin{equation}
 \begin{array}{c}
g=2\,dY\,dQ-\tfrac23F_Q\,dY\,dP+(\tfrac13\Dd F_Q-\tfrac29F_Q^2-F_P)\,dY^2-dP^2,\\
\ \\
\omega=\tfrac23(F_{QP}-\Dd F_{QQ})\,dY+\tfrac23F_{QQ}dP,
\end{array}
\label{Cartan}
\end{equation}
where in these expressions $X$ is fixed. Both $g$ and $\omega$ depend on
$X$ explicitly, but a change in $X$ corresponds to a gauge transformation
of the form (\ref{gauge_ew}). Thus, as long as $W=C=0$, the resulting 
Einstein-Weyl structure is independent of $X$. 

In this approach the EW equations,  $W= C=0$, constitute an overdetermined system of two PDEs for a scalar function $F$ of four variables. One can show that this system is compatible (formally integrable), which follows from the vanishing
of the Mayer bracket $[W,C]=0$ \cite{KruglikovLychagin}.
In other words, this system of third- and second-order PDEs is  in involution (after three prolongations).
The characteristic variety is a complete intersection,  so the general solution is
parametrized by $6=3\cdot2$ functions of 2 variables (we refer to \cite{Cartan2,BCG3,KL2} for the general dimension theory
of solution spaces).
However, the system is point invariant,
so the diffeomorphism freedom is 2 functions of 2 variables, and henceforth the actual
solution space is parametrized by $4=6-2$ functions of 2 variables.
This re-proves  Cartan's count.

\medskip

\noindent

{\bf Proof of Theorem \ref{theo_ms}}.
Setting $A=-\frac{2}{3}F_Q, \ B=\frac{1}{3}\Dd F_Q-\frac{2}{9}F_Q^2-F_P$ one can rewrite (\ref{Cartan}) in the form
$$
 \begin{array}{c}
g=2\,dY\,dQ+A\,dY\,dP+B\,dY^2-dP^2,\\
\ \\
\omega=(A_P-2B_Q-\frac{1}{2}AA_Q)\,dY-A_QdP.
\end{array}
$$
To bring the corresponding EW equations to the desired form we 
fix $X=0$, and set 
the variables as follows:
$Q(0)=x,\ P(0)=y,\ Y(0)=2t, \ A|_{X=0}=a,\ B|_{X=0}= -b-\frac{1}{4}a^2$,
where now $(x, y, t)$ are local coordinates on $M^3$, and 
$a, b:M^3\rightarrow\R$.
This results in
\begin{equation}
\begin{array}{c}
g=4dt\,dx+2a\,dt\,dy-(a^2+4b)\,dt^2-dy^2,\\
\ \\
\omega=(aa_x+2a_y+4b_x)\,dt-a_xdy.
\end{array}
\label{MS1}
\end{equation}
The EW equations reduce to a pair of second-order  conservative PDEs,
 \begin{equation}
(a_t+aa_y+ba_x)_x=(a_y)_y,\qquad
(b_t+bb_x-ab_y)_x=(b_y-2ab_x)_y,
\label{MS2}
\end{equation}
which coincide with the Manakov-Santini system (\ref{MS}) upon substitution $a=v_x, \ b=u-v_y$, see also \cite{Pavlov}.
Note that system (\ref{MS2}) allows one to uniquely reconstruct $g$ and $\omega$ in (\ref{MS1}): the conformal structure $g$ comes from the principal symbol of  system (\ref{MS2}), and $\omega$ is given by formula (\ref{omega}).
Since the construction directly follows from Cartan's approach, we can conclude that the Manakov-Santini system gives {\it all}  EW structures.  
The general solution of   system (\ref{MS2})  depends on 4 arbitrary functions of 2 variables which agrees with Cartan's result.

\koniec

The  Lax representation of (\ref{MS2}) has the form $[X_1, X_2]=0$, where
 $$
X_1=\p_t-(\lambda^2-a\lambda-b)\,\p_x +m\,\p_\lambda,\qquad
X_2=\p_y-\lambda\,\p_x+n\,\p_\lambda,
 $$
and
 $$
m= -a_x\lambda^2+(aa_x-a_y-b_x)\lambda+(ab_x-b_y),\qquad
n= -a_x\lambda-b_x.
 $$
We point out that this Lax pair transforms to the one of the Manakov-Santini system presented in the Introduction via the change of variables  $a=v_x, \ b=u-v_y, \ \lambda=\tilde \lambda+v_x$.  Taking a linear transformation of the Lax
vector fields results in a Lax pair linear in the parameter $\lambda$. A further
affine translation of $\lambda$ with non--constant coefficients can be used
to bring the Lax pair to the canonical form used in \cite{Dun4, Dun3}.

Projecting  integral surfaces of the distribution spanned by $X_1$ and $X_2$ in the 4D space with coordinates
$(x, y, t, \lambda)$ to the space $M^3$ with coordinates $(x, y, t)$ one obtains a 2-parameter family
of null totally geodesic surfaces of the corresponding EW structure. There is an $\RP^1$--worth of such surfaces through any point in $M^3$.

The constraint $a=0, \ b=u$ reduces system (\ref{MS2}) to the dispersionless KP equation, $(u_t+uu_x)_x=u_{yy}$, while the corresponding EW structure reduces to the one from \cite{Dun4}:
$$
g=4dt\,dx-4u\,dt^2-dy^2, ~~~
\omega=4u_x\,dt.
$$ 
Any EW structure which admits a parallel vector field can locally be put in this form. Another possible reduction is $u=0$. This corresponds to the most general hyper-CR Einstein--Weyl structure \cite{Dun7}.

\subsubsection {Translationally non-invariant version of the Manakov-Santini system}

Here our starting point is the general ansatz 
for a metric in the conformal class and a covector \cite{Dun3}.
Using the diffeomorphism and conformal freedom, the representative metric
can be put in the form (\ref{MS1}). Set
$
\omega=\omega_1\,dt+\omega_2\, dx+\omega_3\,dy.
$
Imposing the Einstein-Weyl conditions we obtain a system of 5 PDEs for $a, b, \omega_i$, that is not presented here due to its complexity.
The corresponding Lax pair has the form $[X_1, X_2]=0$, where
 $$
X_1=\p_t-(\lambda^2-a\lambda-b)\,\p_x +m\,\p_\lambda,\qquad
X_2=\p_y-\lambda\,\p_x+n\,\p_\lambda,
 $$
and $m$ and $n$ are the following cubic and quadratic polynomials in $\lambda$:
 $$
m= -\frac{1}{2}\omega_2 \lambda^3+\frac{1}{4}(a\omega_2+4\omega_3)\lambda^2-\frac{1}{2}(\omega_1+b\omega_2+2a\omega_3-aa_x-2b_x)\lambda
$$
$$
+\frac{1}{4}(a\omega_1+ab\omega_2+a^2\omega_3-2aa_y-4b_y),
$$
$$
n= -\frac{1}{4}\omega_2 \lambda^2 +\frac{1}{2}(\omega_3-a_x)\lambda -\frac{1}{4}(\omega_1+b\omega_2+a\omega_3-2a_y).
 $$
One of the five EW equations has the simple form $(\omega_2)_x+\omega_2^2/2=0$. This leads to the natural branching:

\smallskip

\noindent {\bf Case 1}: $\omega_2=0$. Up to further elementary integration and changes of variables this case can be reduced to the
form (\ref{MS1}), with the Manakov-Santini system (\ref{MS2}) for $a, b$.

\smallskip

\noindent {\bf Case 2}: $\omega_2=2/x$ (strictly speaking, $\omega_2=2/(x+f(y, t))$, however, $f(y, t)$ can be removed by a transformation $x\to x+f(y, t)$, that preserves the form of the metric after appropriate re-definition of $a$ and $b$). This branch can be viewed as a translationally non-invariant ($x$-dependent) version of the Manakov-Santini system.

\smallskip

In view of Theorem \ref{theo_ms} both branches are equivalent, but we have 
been unable to find a combination of a conformal rescaling and a coordinate
transformation which reduces Case 2 to Case 1.

\subsection{Einstein-Weyl equations via  Bogdanov's system}

The following system was proposed by Bogdanov \cite{Bogdanov}  as a two-component generalization of the dispersionless Toda equation:
$$
(e^{-\phi})_{tt}=m_t\phi_{xy}-m_x\phi_{yt}, ~~~ m_{tt}e^{-\phi}=m_xm_{yt}-m_tm_{xy}.
$$
It possesses a Lax representation $[X_1, X_2]=0$, where
$$
X_1=\partial_x-\left(\lambda+ \frac{m_x}{m_t}\right)\partial_t+\lambda \left(\phi_t \frac{m_x}{m_t}-\phi_x\right)\partial_{\lambda}, ~~~
X_2=\partial_y+\frac{1}{\lambda} \frac{e^{-\phi}}{m_t}\partial_t+ \frac{(e^{-\phi})_t}{m_t}\partial_{\lambda}.
$$
It was observed in \cite{FerKrug} that, for any solution of the Bogdanov system, the metric
$$
g=(m_xdx+m_tdt)^2+4e^{-\phi}m_tdxdy
$$
and the  covector
$$
\omega=\left(\frac{m_{tt}}{m_t^{2}}-2\frac{\phi_t}{m_t}\right)(m_x\,dx+m_t\,dt)+ 2\frac{m_{yt}}{m_t}\,dy
 $$
satisfy the EW equations. Note that $g$ comes from the principal symbol of the system, and $\omega$ is given by formula (\ref{omega}). The general solution of the Bogdanov system depends on 4 arbitrary functions of 2 variables. It is natural to expect that this gives  (locally)  a {\it generic}  EW structure.

Setting $m=t$ one obtains the $SU(\infty)$ Toda equation \cite{BF}, $(e^{-\phi})_{tt}=\phi_{xy}$, while the corresponding EW structure reduces to the one from \cite{Ward}:
$$
g=dt^2+4e^{-\phi}dxdy, ~~~
\omega=-2{\phi_t}dt.
 $$

\subsection{Einstein-Weyl equations in diagonal coordinates}

Note that any three-dimensional metric possesses diagonal coordinates depending locally on 3 arbitrary functions of 2 variables \cite{Cartan2, DeTurck}. We can therefore use  conformal freedom
$g\rightarrow \varphi g,\  \omega\rightarrow \omega+d\ln\varphi$ to set
$$
g=a^2dt^2-dx^2+b^2dy^2,\qquad \omega=\omega_1dx+ \omega_2dy+ \omega_3dt.
$$
In this case the EW equations give rise to a system of five PDEs for the five functions
$a, b, \omega_i,$ which are second-order in $a, b$, and first-order in $\omega_i$ (this system is not presented explicitly due to its complexity).
It possesses the Lax pair $[X_1, X_2]=0$:
$$
X_1=\partial_t-a\cos \lambda \ \partial_x+m\ \partial_{\lambda}, ~~~
X_2=\partial_y-b\sin \lambda \ \partial_x+n\ \partial_{\lambda},
$$
where
\begin{eqnarray*}
m&=& -\frac{a}{2b}\omega_2\sin^2\lambda-\frac{1}{2}\omega_3\sin\lambda\cos\lambda
+(\frac{1}{2}a\omega_1-a_x)\sin\lambda+\frac{a_y}{b}, \\
n&=& \tfrac{b}{2a}\omega_3\cos^2\lambda+\tfrac{1}{2}\omega_2\sin\lambda\cos\lambda
-(\tfrac{1}{2}b\omega_1-b_x)\cos\lambda-\tfrac{b_t}{a}.
\end{eqnarray*}
The general solution of the  system for  $a, b, \omega_i$ depends locally
on $7=2\cdot2+3\cdot1$ arbitrary functions of 2 variables  (recall the order of PDEs).
Since diagonal coordinates exist with the freedom of 3
arbitrary functions of 2 variables, this again confirms that EW structures depend
on $4=7-3$ arbitrary functions of 2 variables.

The above system  possesses a reduction \cite{FerKrug}
$$
g=(1-e^{-u})dt^2-dx^2+(e^u-1)dy^2, ~~~
\omega=\frac{e^{u}+1}{e^{u}-1}u_xdx-u_ydy+u_tdt,
$$
for which the EW equations reduce to the scalar second-order PDE
$$
u_{xx}+u_{yy}-(\ln(e^u-1))_{yy}-(\ln(e^u-1))_{tt}=0.
$$
This  is  the dispersionless limit of the `gauge-invariant'  Hirota equation \cite{FNR}. 

\section{Anti-self-duality equations}

A conformal structure $g$ on a four-dimensional manifold is called anti-self-dual (ASD)
if the self-dual (SD) part of its conformal Weyl tensor vanishes: $W_+=\frac12(W+*W)=0$. The twistor-theoretic integrability of the ASD condition was established in \cite{Penrose}. It was shown in \cite{Grossman} that generic ASD structure depends on 6 arbitrary functions of 3 variables. The existence of a Lax pair is implicitly built in the fact that any ASD structure possesses a 3-parameter family of totally null $\alpha$-surfaces. Below we present several  forms of ASD equations and their  Lax pairs in  specially adapted coordinates.


\subsection{Anti-self-duality equations in Pleba\'nski-Robinson coordinates}

Here we present  explicit formulae, including the corresponding Lax pair, in Pleba\'nski-Robinson coordinates  $(w, z, x, y)$ where
the metric $g$ in the ASD conformal class on an open set $M^4\subset\R^4$
takes the hyper-heavenly form,
\be
\label{ASDmetric}
g=dwdx+dzdy+pdw^2+2qdwdz+rdz^2,
\ee
where $p, q, r$ are functions of all four variables. We  assume that all coordinates are real, so that the signature of $g$ is $(2, 2)$.
\begin{prop}
\label{propo_ASD}
The metric (\ref{ASDmetric}) has ASD Weyl tensor
if the functions  $p, q, r$ satisfy the system of three second order PDEs :
 \begin{equation}
\begin{array}{c}
p_{xx}+2q_{xy}+r_{yy}=0,\\
\ \\
m_x+n_y=0,\\
\ \\
m_z-qm_x-rm_y+(q_x+r_y)m=n_w-pn_x-qn_y+(p_x+q_y)n,
\end{array}
\label{SD}
 \end{equation}
where
 \begin{equation}\label{mnref}
m:= p_z-q_w+pq_x-qp_x+qq_y-rp_y, ~~~
n:=q_z-r_w+qr_y-rq_y+pr_x-qq_x.
 \end{equation}
Conversely, any ASD conformal structure is locally of the form (\ref{ASDmetric}),
where $(p, q, r)$ satisfy   system (\ref{SD}).
\end{prop}

\noindent {\bf Proof.}  We will make use of the isomorphisms
${\Lambda^2}_+=S'\odot S'$ and  $TM^4=S\otimes S'$, where
$S$ and $S'$ are real rank-two symplectic vector bundles (spin bundles), 
${\Lambda^2}_+$ is the rank-three bundle of self-dual two-forms on $M^4$
and $\odot$ is the symmetrized tensor product.
The seminal result of Penrose \cite{Penrose} asserts that the
ASD condition is equivalent to the existence of a 3-parameter family of $\alpha$-surfaces
(totally null surfaces in $M^4$ with SD tangent bi-vector). This means that any section of $S'$ corresponds
to a SD two-form defining a 2D distribution integrable in the Frobenius sense.
To arrive at the canonical form  (\ref{ASDmetric}) we select a 2-parameter family of $\alpha$--surfaces
corresponding to a section $\iota\in\Gamma(S')$.
Let $\Sigma\in\Gamma({\Lambda^2}_+)$ be a SD two--form corresponding to this section.
It is Frobenius-integrable so there exist independent functions $x$ and $y$ on $M^4$ such that
$\mbox{Ker}(\Sigma)=\mbox{Span}(\p/\p x, \p/\p y)$.
We can moreover rescale the spinor $\iota$ so that the corresponding two--form $\Sigma$ is closed and proportional to  $dw\wedge dz$.
Therefore $w$ and $z$ are constant on each $\alpha$--surface in the 2-parameter family, and $(x, y)$
are coordinates on the surface. The $\alpha$--surfaces are totally null so that the conformal structure
is represented by
\be
\label{metric_tetrad}
g={\bf e}^{00'} {\bf e}^{11'}-{\bf e}^{01'} {\bf e}^{10'},
\ee
where
\be
\label{one_forms}
{\bf e}^{00'}=adz,\quad
{\bf e}^{10'}=bdw,\quad {\bf e}^{01'}=-dx-pdw-qdz,
\quad
{\bf e}^{11'}=
dy+qdw+rdz;
\ee
here $(a, b,  p, q, r)$ are so far unspecified functions
(we have set the $dz dx$ and $dw dy$  coefficients in $g$ to $0$ by exploiting
the coordinate freedom in the choice of $(x, y)$).
To examine the ASD condition we choose a basis of $S'$ consisting of
two spinors $(o, \iota)$. The self-dual Weyl spinor $W'$ is a section
of $\mbox{Sym}^4(S')$ given by
 \[
W'=W_0\; o\;o\;o\;o + 4W_1 \;o\;o\;o\;\iota +6W_2\; o\;o\;\iota\;\iota+ 
4W_3 \;o\; \iota\;\iota\; \iota+W_4\;\iota\;\iota\;\iota\;\iota,
\]
where the symmetrised tensor product is implicit in this formula.
We find that $W_4$ vanishes identically and that
$W_3=\frac14 \p_x\p_y \mbox{ln}(a/b)$.  Therefore
$a=b\exp{(\alpha+\beta)}$, where
$\alpha=\alpha(x, w, z), \beta=\beta(y, w, z)$.

Now we make a coordinate transformation
$x\rightarrow \tilde{x}(x, w, z), y\rightarrow \tilde{y}(y, w, z)$ such that
$\p\tilde{y}/\p y=\exp{(-\beta)}$ and $\p\tilde{x}/\p x=\exp{(\alpha)}$.
Finally we redefine $(p, q, r)$ and conformally rescale the resulting
metric by $b^{-1}\exp{(-\alpha)}$. This puts the metric in the form
(\ref{ASDmetric}), with the corresponding null tetrad given by
(\ref{one_forms}) with $a=b=1.$
So far our proof has more or less followed the construction of Pleba\'nski
and Robinson \cite{PR}, but now we shall proceed differently. Instead of imposing the Einstein equations we shall assume that the remaining three
components of $W'$ vanish. This gives the coupled system
(\ref{SD}). \qed

\bigskip

System (\ref{SD}) possesses the Lax pair $[X_1, X_2]=0$, where $X_1$ and $X_2$ are $\lambda$-dependent vector fields,
 \begin{eqnarray}
\label{ASDLax}
X_1&=&\partial_w-p\partial_x-q\partial_y+\lambda\p_y+[m-\lambda(p_x+q_y)] \partial_{\lambda},\\\nonumber
X_2&=&\partial_z-q\partial_x-r\partial_y-\lambda\p_x+[n-\lambda(q_x+r_y)] \partial_{\lambda},
 \end{eqnarray}
where $m,n$ are given by expressions (\ref{mnref}).
This Lax pair is a coordinate realization of the general twistor distribution $L_A=(X_1, X_2)$ on the projectivized spin bundle $S'$ given by
\be
\label{twistor_lax}
L_A=\pi^{A'}{\bf e}_{AA'}-\pi^{A'}\pi^{B'}\pi^{C'}\Gamma_{AA'B'C'}\frac{\p}{\p\lambda},
\ee
where the indices $A, B, A', B', \dots$ take values $0$ or $1$, the vector fields
${\bf e}_{AA'}$ are dual to the one forms (\ref{one_forms}), $\Gamma_{AA'B'C'}$ are components of the spin connection, and $\pi^{A'}=(1, \lambda)$
are homogeneous coordinates on the fibres of $\mathbb{P}{S'}$.
Projecting integral surfaces of the distribution spanned by $X_1$ and $X_2$ from the correspondence space
${M}^4\times\RP^1$  with coordinates $(w, z, x, y,\lambda)$ to  $M^4$,
we obtain a 3-parameter family of null  surfaces ($\alpha$-surfaces) of the  conformal structure $g$. 
The spectral parameter $\lambda$ on $\RP^1$ is a coordinate on the circle of 
$\alpha$--surfaces
at each point of $M^4$.
The conformal structure (\ref{ASDmetric}) can be read off the principal symbol of system (\ref{SD}). Indeed,  the principal symbol of (\ref{SD}) equals $Q^3$, where
 $$
Q=\p_w\p_x+\p_z\p_y- p\,\p_x^2-2q\,\p_x\p_y-r\,\p_y^2,
 $$
and the inverse matrix  of the symmetric bivector $Q$ defines  conformal structure (\ref{ASDmetric}).

Theorem (\ref{prop_ASD}) states that a further simplification is possible, 
so that  ASD conditions reduce to a system of 2 third-order PDEs for 2 
functions.
The proof below  uses one of the equations from Proposition 
(\ref{propo_ASD}) 
as integrability conditions.

\medskip

\noindent {\bf Proof of Theorem \ref{prop_ASD}.} Rewrite the first equation in (\ref{SD}) as $(p_x+q_y)_x+(q_x+r_y)_y=0$, which implies the existence of a function
$s$ such that $p_x=s_y-q_y$ and $r_y=-s_x-q_x$. These two equations can again be regarded
as the integrability conditions for the existence of two functions
$F$, and $G$ on $M^4$ such that
\[
p=F_y,\quad \ q=-(F_x+G_y)/2,\quad \ r=G_x.
\]
The remaining two equations in (\ref{SD}) now yield (\ref{sd_3rd}).

To exhibit a simple Lax pair for (\ref{sd_3rd})  we shall
make a linear transformation (null rotation) of the frame of $S'$ which does not change  metric (\ref{metric_tetrad}):
\[
{\bf e}^{11'}\rightarrow {\bf e}^{11'}+\gamma {\bf e}^{10'}=dy-G_ydw+G_xdz, \quad
{\bf e}^{01'}\rightarrow {\bf e}^{01'}+\gamma {\bf e}^{00'}=
-dx-F_ydw+F_xdz,
\]
here $\gamma=(F_x-G_y)/2$. In this spin frame,  Lax pair (\ref{twistor_lax})  gives (\ref{new_lax}).
\qed

\subsection{Anti-self-duality equations and torsion-free ODE systems}

In the spirit of Cartan, it was shown by  Grossman \cite{Grossman} that there is a one-to-one correspondence between ASD conformal structures in signature $(2, 2)$  and systems of second-order ODEs with vanishing generalized Wilczynski invariants (torsion-free systems in his terminology). In particular, Grossman has  shown
that a generic torsion-free system depends on 6 arbitrary functions of 3 variables. The canonical form (\ref{ASDmetric1}) of the ASD metric can be directly derived from  Grossman's approach \cite{CDT12}: if a torsion-free system of 2 ODEs is of the form
\[
W''=G(X, W, Z, W', Z'), \quad Z''=F(X, W, Z, W', Z'),
\]
(here prime denotes differentiation by $X$), then the solution space $M^4$ can be parametrized by 
fixing $X$, say $X=0$, and defining $(w, z, x, y)$ to be the initial conditions:  
$w=W(0),\ z=Z(0),\ y=W'(0),\ x=-Z'(0)$.
The conformal ASD structure on $M^4$ is then defined by demanding that
 points in the $(X, W, Z)$ space correspond to totally null $\alpha$-surfaces
in $M^4$. In the chosen coordinates this leads to the formula
(\ref{ASDmetric1}), where $F, G$ are evaluated at $X=0$, see  \cite{CDT12} for details of this construction.

\subsection{Anti-self-duality equations in doubly biorthogonal coordinates}

It was demonstrated in \cite{Grant1} that any (analytic) four-dimensional metric can be brought into  block-diagonal form,
 $$
g=\left(
\begin{array}{cccc}
a_1&a_2&0&0\\
a_2&a_3&0&0\\
0&0&b_1&b_2\\
0&0&b_2&b_3
\end{array}
\right).
 $$
   Coordinates of this type are known as doubly biorthogonal.  They depend locally on 4 arbitrary functions  of 3 variables. Using the conformal freedom to set  $\det g=1$, one can show that the equations of
  self-duality reduce to a (complicated) system of 5 second-order PDEs for the 5 (independent)
  functions among $a_i$ and $b_i$. The general solution of this  system depends locally on 10 arbitrary
  functions of 3 variables. Since biorthogonal coordinates exist with the freedom of 4 arbitrary
  functions of 3 variables, this is again in agreement with the fact that  self-dual structures depend
  locally on $10-4=6$ arbitrary functions of 3 variables.

\subsection{Reductions of self-duality equations}

 ASD equations   possess several geometric reductions of interest:

\medskip

\noindent {\bf Hyper-Hermitian case:}
this case is characterized by the existence of a Lax pair that does not contain  derivatives with respect to the spectral
parameter \cite{Dun5,Calderbank}. Taking into account (\ref{new_lax}) this leads to a pair of second-order PDEs
\be
\label{HHS}
Q(F)=0,\quad Q(G)=0.
\ee
This system was first  derived in \cite{FP},
where
the corresponding conformal structures were referred to as  `weak heavenly spaces'.
The dispersionless Lax pair for  system (\ref{HHS}) is
\[
X_1=\p_w-F_y\p_x+G_y\p_y+\lambda\p_y, \quad X_2=\p_z+F_x\p_x-G_x\p_y-\lambda\p_x.
\]
In \cite{Dun5} it was shown that all (pseudo) hyper-Hermitian conformal structures locally arise from
solutions to (\ref{HHS}). The general solution to this system depends on 4 arbitrary functions of 3 variables.
In the special case where $F=\theta_y,  G=\theta_x$, the hyper-Hermitian
system  reduces to Plebanski's 2nd heavenly equation,
\be
\label{heavenly}
\theta_{yz}+\theta_{xw}+\theta_{xy}^2-\theta_{xx}\theta_{yy}=0,
\ee
and the metric $g$ is Ricci--flat. It depends on 2 arbitrary functions of 3 variables.

\medskip

\noindent {\bf Null K\"ahler case:}
the ansatz $F=\theta_y, G=\theta_x$
reduces  ASD equations   (\ref{sd_3rd})
to a single fourth-order PDE for $\theta$ \cite{DP}:
\begin{equation}
 \begin{array}{c}
\quad Q(f)=0, ~~~ f=\theta_{yz}+\theta_{xw}+\theta_{xy}^2-\theta_{xx}\theta_{yy},\\
\ \\
\mbox{where}\quad Q= \p_w\p_x+\p_z\p_y-\theta_{yy}{\p_x}^2-\theta_{xx}{\p_y}^2+2\theta_{xy}\p_x\p_y.
\end{array}
\label{NKS}
\end{equation}
In this case the self-dual two form $\Sigma=dw\wedge dz$ corresponding to the two-parameter family of $\alpha$--surfaces from the proof of Theorem 3.1
is covariantly constant. Conversely, it was demonstrated in \cite{DP} that any ASD metric $g$ that admits a self-dual covariantly constant two-form $\Sigma$
such that $\Sigma\wedge\Sigma=0$, is locally given by a solution to (\ref{NKS}).
The dispersionless Lax pair for (\ref{NKS}) is
 $$
X_1=\partial_w-\theta_{yy}\partial_x+\theta_{xy}\p_y+\lambda\partial_y+f_y\partial_{\lambda},
 $$
 $$
X_2=\partial_z+\theta_{xy}\partial_x-\theta_{xx}\partial_y-\lambda\p_x-f_x\partial_{\lambda}.
 $$
In the special case $f=0$ we recover the second heavenly equation (\ref{heavenly}).
\medskip

\noindent {\bf Other reductions:}
The coordinate system introduced in Proposition
\ref{propo_ASD}
is adapted to a choice of a preferred two--parameter family of
$\alpha$--surfaces determined by a section $\iota\in\Gamma(S')$,
or equivalently by 
 a Frobenius-integrable simple two form $\Sigma$.
There are other possibilities 
which single out a non-degenerate
two form $\Sigma$ such that $\Sigma\wedge\Sigma\neq 0$.
This requires a choice of two independent sections of $S'$ and
leads to PDEs generalising Plebanski's 1st heavenly equation
\cite{Plebanski, BP}. In particular, the Przanowski equation \cite{Przanowski}
describing all ASD Einstein metrics with non-vanishing cosmological
constant, is written down in such coordinates.
A Lax pair for this 
equation has recently been found in \cite{Hoegner}.
Its 2nd heavenly form analogous to (\ref{heavenly}) has been given in 
\cite{Chudecki}.


\end{document}